\documentclass[10pt,two column,aps]{revtex4-1}
\usepackage{xcolor}
\usepackage[utf8]{inputenc}

 \usepackage{fancyhdr} 

\usepackage{amsmath}
\usepackage{amsfonts}
\usepackage{amssymb}
\usepackage{graphicx,hyperref}
\usepackage{amssymb}
\usepackage{mathptmx}

\usepackage{cleveref}
\newcommand{\be}{\begin {equation}}
\newcommand{\ee}{\end {equation}}

\newcommand{\mb}{\mathbf}
\hypersetup{
    colorlinks,    citecolor=blue,    filecolor=black,    linkcolor=blue,    urlcolor=blue
}

\newcommand{\ptdd}[1]{\frac{\partial^2 #1}{\partial t^2}}

\newcommand{\psdd}[1]{\frac{\partial^2 #1}{\partial z^2}}

\begin{document}

\title{\Large{Chirp assisted ion acceleration via relativistic self induced transparency}}
\author{Shivani Choudhary}
\email{shivani.choudhary@pilani.bits-pilani.ac.in}
\author{Amol R. Holkundkar}
\email{amol.holkundkar@pilani.bits-pilani.ac.in}

\affiliation{Department of Physics, Birla Institute of Technology and Science - Pilani, Rajasthan, 333031, India}

\begin{abstract}
We study the effect of the chirped laser pulse on the transmission and associated ion acceleration by the sub-wavelength target. In the chirped laser pulses, the pulse frequency has a temporal variation about its fundamental frequency, 
  which manifests to the temporal dependence of the critical density ($n_c$). In this work we used a chirp model which is beyond the linear approximation. For negatively (positively) chirped pulses, the high (low) frequency component of the pulse interacts with the target initially followed by the low (high) frequency component. The threshold plasma density for the transmission of the pulse is found to be higher for the negatively chirped laser pulses as compared to the unchirped or positively chirped pulses. The enhanced transmission of the negatively chirped pulses for higher densities ($6n_c$) results in very efficient heating of the target electrons, creating a very stable and persistent longitudinal electrostatic field behind the target. The void of the electrons results in expansion of the target ions in either direction, resulting in the broad energy spectrum. We have introduced a very thin, low density ($< n_c$) secondary layer behind the primary layer. The ions from the secondary layer are then found to be accelerated as a mono-energetic bunch under the influence of the electrostatic field created by the primary layer upon interaction by the negatively chirped pulse. Under the optimum conditions, the maximum energy of the protons are found to be $\sim 100$ MeV for 10 fs (intensity fwhm); Circularly Polarized; Gaussian; negatively chirped laser pulse with peak intensity $\sim 8.5\times 10^{20}$ W/cm$^2$. 
    
\end{abstract}

\date{\today}

\maketitle

\section{Introduction}

The advent of high power lasers promised the vast number of applications, covering both applied and fundamental aspects of basic sciences. The laser-plasma based acceleration of the ions and electrons paved the possibility of constructing a table-top \cite{Mangles2004_Nature,Faure2004_Nature,Geddes2004_Nature}, high energy, charged particle beams for medical \cite{Karsch2017_TandF,Bulanov2014_PU} and industrial applications \cite{Cobble2002_JAP,Barberio2017_nature}. In the last couple of decades, we have already witnessed the experimental realizations of the acceleration of ions to multi MeV of energies via various acceleration mechanisms. The Target Normal Sheath Acceleration (TNSA) \cite{Snavely2000_PRL,Passoni2010_NJP}, Radiation Pressure Acceleration (RPA) \cite{Qiao2011_POP,Scullion2017_PRL}, Hole Boring (HB) \cite{Robinson2012_PPCF}, Breakout Afterburner (BOA) \cite{yin2011_POP}, are among few very well studied mechanisms both theoretically and experimentally. A wonderful review of this contemporary field of laser-induced ion acceleration is presented by Macchi et al. \cite{Macchi2013_RMP} in which all the ion acceleration mechanisms are covered in depth.

The laser-driven ion acceleration in the relativistic self-induced transparency (RSIT) regime is also proving to be the fascinating mechanism to have a very efficient high energy ion and neutron beams \cite{Willingale2009_PRL,Roth2013_PRL,henig2009_PRL,Jiao2017_SD,Higginson2018_nature}. The ion and neutron beams with energies around $\sim 180$ MeV have been observed in experiments by the RSIT mechanism \cite{henig2009_PRL,Roth2013_PRL}. In general, the dispersion relation for the electromagnetic wave propagation in plasmas, $\omega^2 = k^2c^2 + \omega_p^2$, ignores the interaction of the laser fields with the medium, which in principle alters the electron density and so the dispersion relation via the plasma frequency $\omega_p = \sqrt{n_e e^2/\varepsilon_0m_e}$. The critical plasma density, $n_c = \varepsilon_0 m_e \omega^2/e^2$, as predicted by the above dispersion relation has to be corrected for the laser-plasma interaction dynamics. If we take into account the fact, that the plasma electrons will respond to the laser electric field, the modified dispersion relation reads, $\omega^2 = k^2c^2 + \omega_p^2/\gamma$, where $\gamma = \sqrt{1 + (\mb{p}/m_e c)^2}$ and $\mb{p}$ is the electron momentum. The high-intensity laser beams can efficiently heat the plasma electrons, as a consequence, the ion acceleration can be enhanced  \cite{Eremin2010_POP,Willingale2009_PRL,Jung2015_LPB,Sahai2013_PRE,Hegelich2013_NJP,Juan2017_POP, Poole2018_iop}. 
 
The introduction of the chirp in the laser pulse is also proving to be a promising way to have enhanced ion energy beams \cite{Vosoughian2015_POP,Mackenroth2016_PRL,Salamin2012_PRA,Holkundkar2015_POP,Vosoughian2017_POP, Souri2018_POP}. In the chirped pulses, the pulse frequency has temporal variation about its fundamental frequency, which manifests in the temporal dependence of the critical density $n_c$ as well. In this study, we aim to characterize the effect of laser pulse chirp on the ion energies under RSIT regime. We consider a chirp model which is beyond the linear approximation \cite{Mackenroth2016_PRL}, the chirp model used is in close analogy with the experimental technique for the pulse amplification i.e. Chirped Pulse Amplification. In order to understand how the chirp of the laser pulse affects the transmission through the target, we developed a simplified wave propagation model for the laser with $a_0 < 1$. The model takes into account the chirp of the pulse while calculating the target density as pulse traverses the target. The results of this simplified wave propagation are found to be consistent with the 1D PIC simulation. Furthermore, we consider the dual layer sub-wavelength target to have a very efficient generation of the accelerated ion bunch from the secondary layer \cite{Gonzalez2016_nature,Scullion2017_PRL,Vosoughian2017_POP}. Recently, the effect of the pulse shape on the enhanced ion acceleration is reported in Ref. \cite{Souri2018_POP} under Radiation Pressure regime of ion acceleration. However, in this study we are concerned in the relativistic transparency regime, and how the chirp of the laser pulse can affect the transmission of the pulse through the target.    

The paper is organized as follows. In Section-II, we discuss the simplified wave propagation model for low laser amplitudes  ($a_0 = 0.5$) and its utility to compute the transmission coefficients for different chirp values.  We compare the results of the wave propagation model with the 1D PIC simulations. In Section-III, we focus on the study of high intense lasers ($a_0 = 20$) with the target. The results showing the effect of the laser pulse chirp on the longitudinal electric field for different target parameters is discussed. Moreover, we consider the need for the second layer to obtain an energetic ion bunch.  The optimization study for different target parameters is then carried out in Section-IV, and finally, we give the summary and concluding remarks in Section-V. 

 \begin{figure}[t]
 \includegraphics[totalheight=7cm]{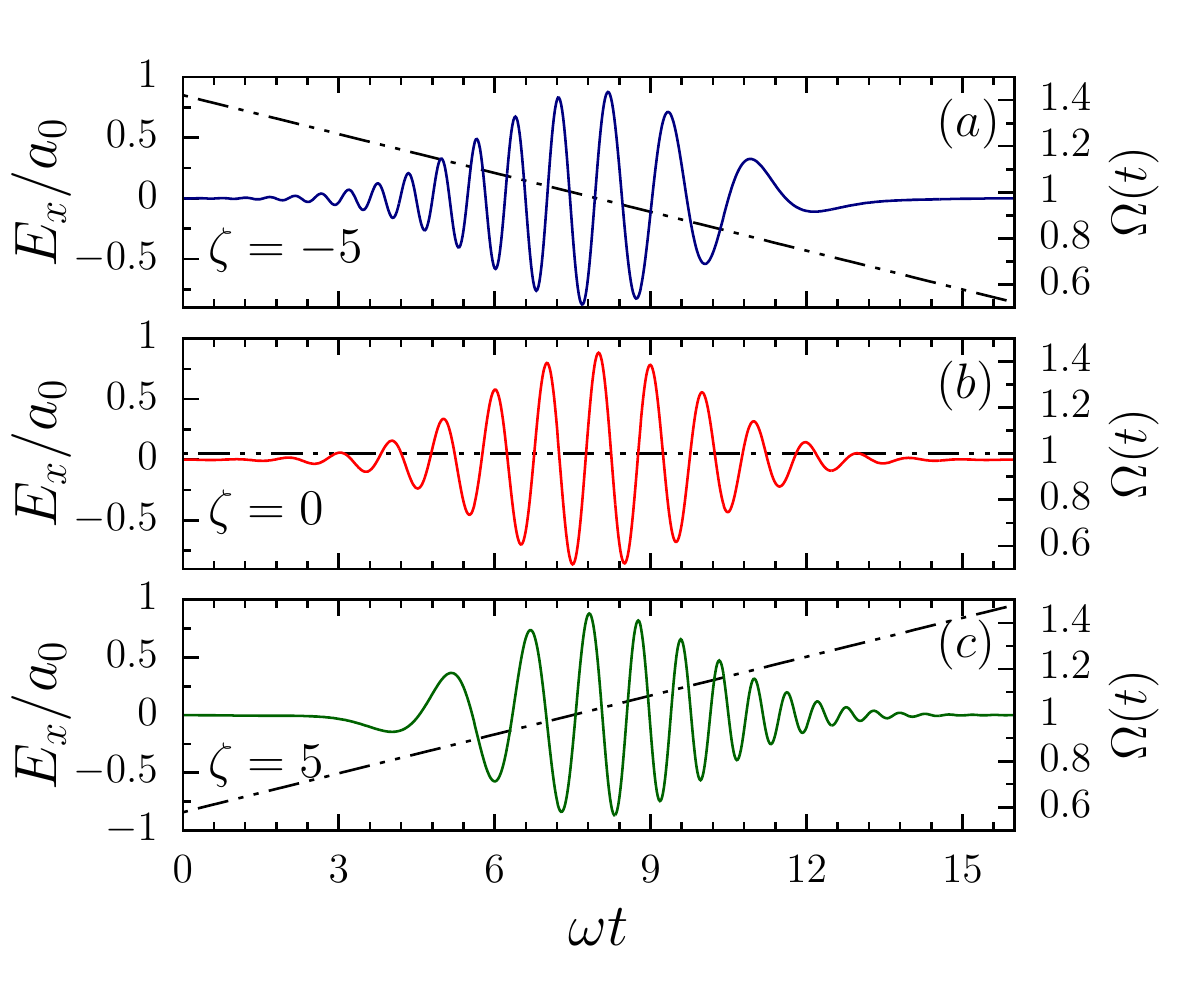}
 \caption{Laser pulse profiles with negative (a), unchirped (b) and positive (c) chirp parameters. Time dependent frequency is also illustrated (dotted line) for each case as well.    }
 \label{pulse}
\end{figure}
  
\section{Theory and simulation model}

The interaction of the intense laser beams with the over-dense plasmas are in general modeled by the cold-relativistic fluid model \cite{Goloviznin2000_POP,Cattani2000_PRE,Tushentsov2001_PRL,Eremin2010_POP,Siminos2012_PRE}. The use of the cold-relativistic fluid model is justified, as the quiver velocity of the electrons involved in laser-plasma interaction exceeds the electron temperatures, and ions are considered as immobile in the time scales of the interests. The threshold plasma densities for RSIT as obtained by seeking the stationary solutions of the cold relativistic fluid model for a semi-infinite plasma slab is reported in Refs. \cite{Cattani2000_PRE,Eremin2010_POP,Siminos2012_PRE}. 

The set of relativistic cold fluid equations are in general challenging to solve for very thin targets, and hence kinetic simulations are routinely used for studying the laser interaction with thin overdense plasma layers  \cite{Shen2001_PRL,Emmanuel2005_POP,Gonoskov2009_PRL,Loch2016_POP}. In this paper, rather than obtaining the stationary solutions for a threshold plasma density, we studied the effect of pulse chirp on the transmission coefficient of the target for a given laser and target parameters. We have used the sub-wavelength target as it allows the transmission from the slightly over-dense plasmas as well. The transmission coefficients are calculated by numerically solving the wave propagation equation along with the corrected electron density, taking into account the time-dependent frequency and amplitude of the circularly polarized chirped laser pulse. In the following, we present our simplified wave propagation model to study the effect of the pulse chirping on the transmission coefficient, followed by the comparison with the 1D PIC simulation.

\begin{figure}[b]
 \includegraphics[width=9cm,height=6cm]{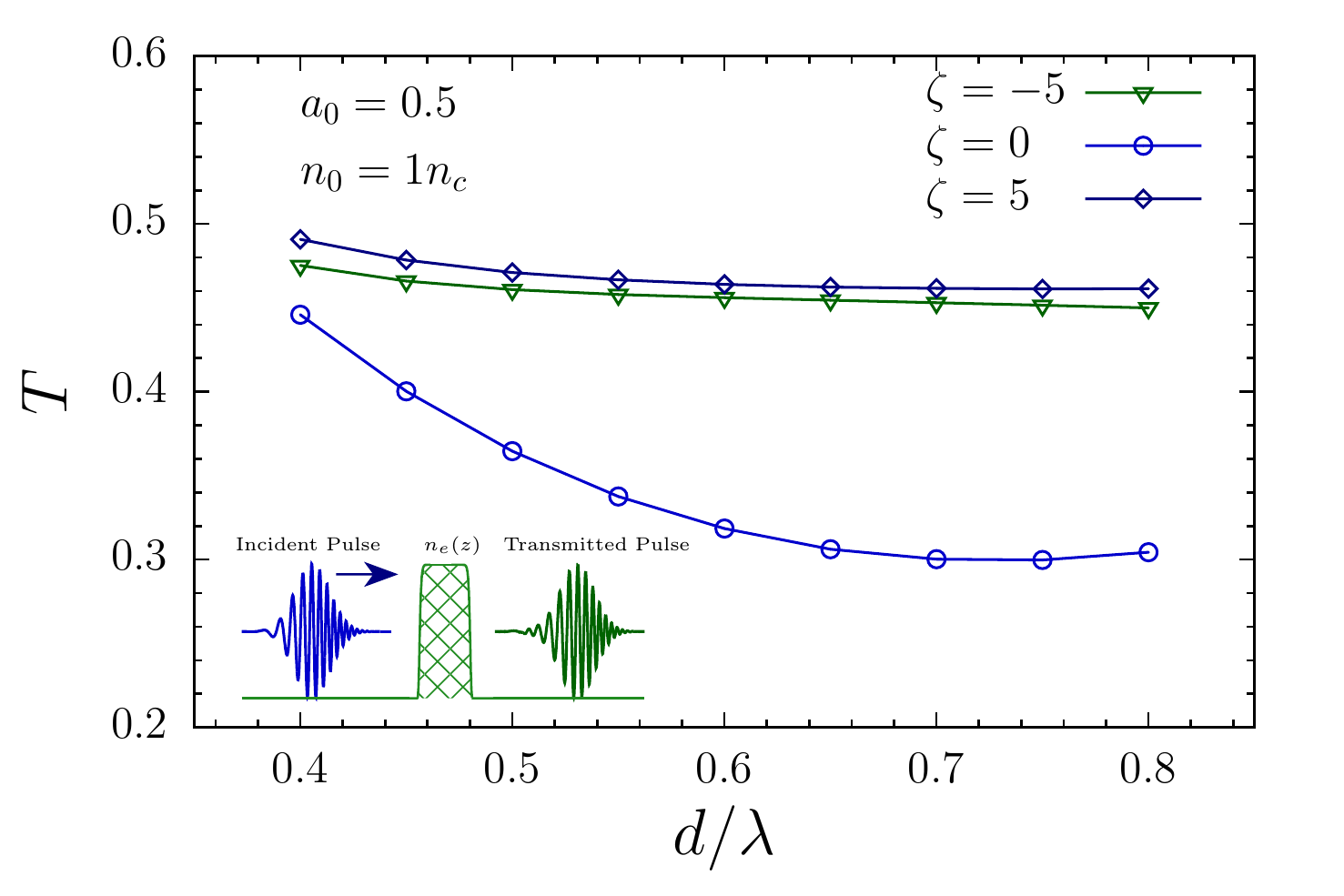}
 \caption{Transmission coefficient of the laser pulse ($a_0 = 0.5$, $\tau = 5$ cycles) for chirp parameters $\zeta = -5,0,5$ for different target thickness of density $n_0 = 1n_c$ is compared. The schematic diagram representing the target geometry (Eq. \ref{nez}), incident pulse and transmitted pulse is illustrated as an inlet. }
 \label{waveTrans}
\end{figure} 

\subsection{Wave propagation model}

Throughout the paper we will be using the dimensionless units. The laser amplitude is normalized as $\mathbf{a}=e\mb{A}/m_{e}c$, where $\mb{A}$ is the vector potential associated with laser, $e$ and $m_e$ are charge and mass of the electron. The time and space are normalized against the laser frequency ($\omega$) and wave number ($\omega t\rightarrow t$ and $kx\rightarrow x$) respectively. The electron density is normalized against the critical density $n_{c}=\varepsilon _{0}\omega ^{2}m_{e}/e^{2}$. In the dimensionless form the EM wave propagation in plasma can be written as \cite{Choudhary2016_EPJD},  
\be  \psdd{\mb{a}} -  \ptdd{\mb{a}} =  \frac{n_e}{\gamma_e} \mb{a} \label{wavepropa}\ee
where, $\gamma_e$ is the relativisitic factor for electron and $n_e$ is the electron density. Using the definition of canonical momentum, the $\gamma_e$ can be expressed as,  
\be  \gamma_e  = \sqrt{1 + \mb{a}^2 + (p_z^e)^2} \label{gama}\ee
here, $p_z^e$ is the dimensionless ($p_z^e/m_e c \rightarrow p_z^e$) longitudinal component of the electron momentum.  In general the electron density will be a function of both $z$ and $t$.  The spatial dependence of the electron density is because of the finite target geometry, however the temporal dependence comes via the chirp of the laser pulse. In a chirped laser pulse the frequency varies with time and hence the associated critical density [$n_{c}(t) =\varepsilon _{0}\Omega(t)^{2}m_{e}/e^{2}$] for the laser pulse will also vary accordingly. It should be noted that, here we are not solving the full set of dynamical fluid equations, and hence the electron density is not going to evolve with time by continuity equation.  In the later part we will see, that this approximation is valid if we intend to calculate the transmission coefficient of the target for the laser pulses with $a_0 < 1$. Furthermore, for $a_0 < 1$ one can ignore the longitudinal electron heating and so Eq. \ref{gama} reduces to, 
\be \gamma_e \sim \sqrt{1 + \mb{a}^2} \label{gama2}.\ee 
The electron density profile in space and time is then given by,  
\be n_e(z,t) \equiv \frac{n_e(z)}{\Omega(t)} = \frac{n_0}{\Omega(t)} \exp\Big[ - 2^{24} \ln(2) \frac{(z - z_0)^{24}}{d^{24}} \Big] \label{nez}\ee
here, $\Omega(t)$ is a time dependent frequency of the chirped laser pulse (at the peak of the laser pulse $\Omega = 1$), $n_0$ is the peak target density, $d$ is its thickness and $z_0$ is the location of the target center. The electron density profile (Eq. \ref{nez}) serves the dual purpose, not only it mimics the thin layer target, rather it is a continuous function of $z$ as well, which is desirable for numerical stability.

 The simulation domain is considered to be $L = 100\lambda$ long, and the target (see Eq. \ref{nez}) of thickness $d$ is placed at 25$\lambda$ ($z_0 = 25\lambda + d/2$). The transmission coefficient ($T$) is calculated by taking the ratio of transmitted pulse energy to the incident pulse energy. The boundary condition on the left of the simulation domain is precisely the temporal profile of the laser pulse. In this paper we have used the laser pulse model as proposed by Mackenroth et. al. \cite{Mackenroth2016_PRL}, because it models the laser pulse chirp beyond the linear approximation \cite{Holkundkar2015_POP, Vosoughian2015_POP,Vosoughian2017_POP,Salamin2012_PRA} and it is also in close analogy to the model of chirped pulse amplification \cite{Mourou2006_RMP}. The boundary conditions on left side of the simulation domain for $\mb{a} = a_x \hat{x} + a_y \hat{y}$ read as, 

 \begin{figure}[t]
 \includegraphics[width=9cm,height=6cm]{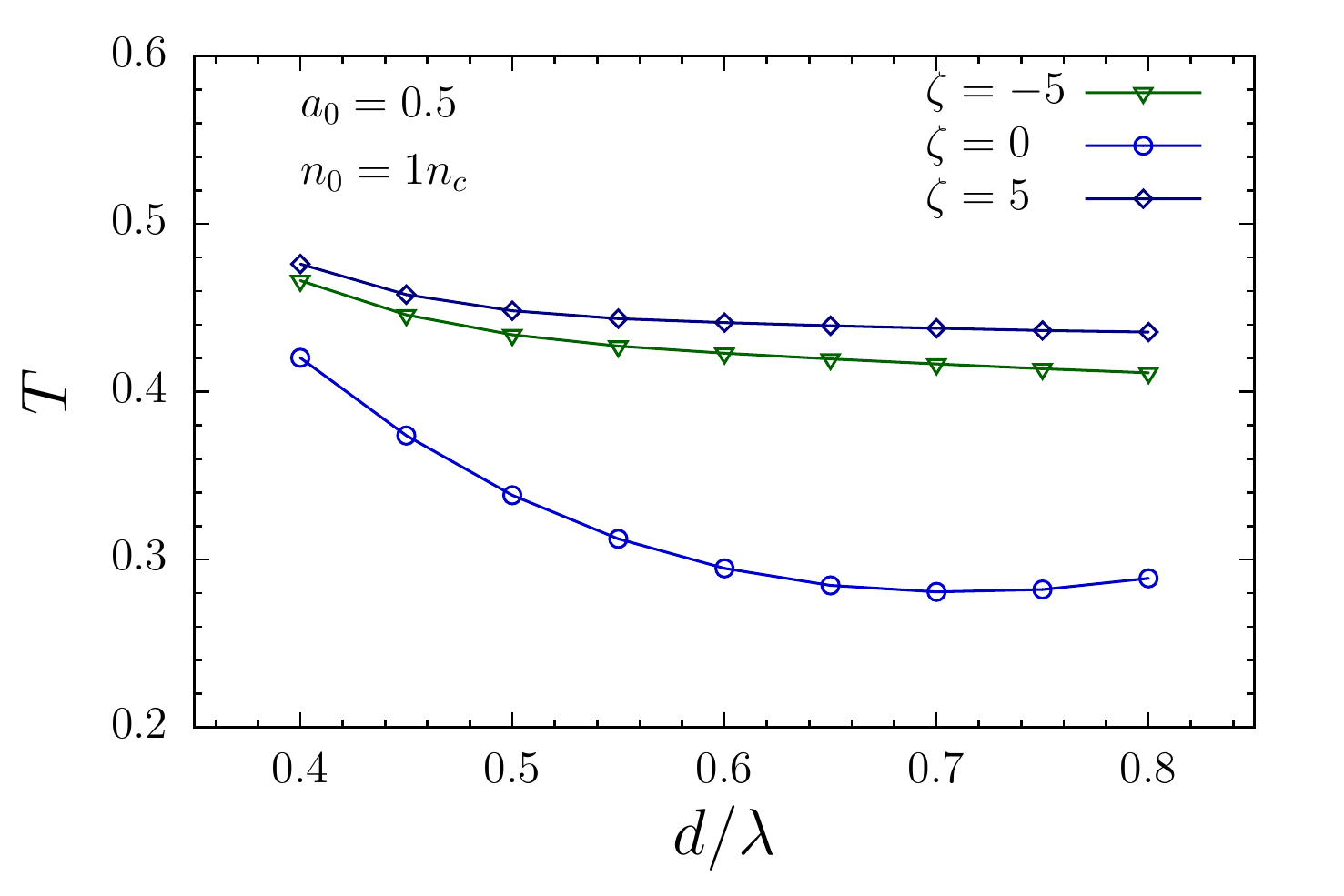}
 \caption{PIC simulation for transmission coefficient of the laser pulse ($a_0 = 0.5$, $\tau = 5$ cycles) for chirp parameters $\zeta = -5,0,5$ for different target thickness of density $n_0 = 1n_c$ is compared.   }
 \label{lpicTrans}
\end{figure}

\be a_x(0,t) = \frac{a_0}{\sqrt{2}} \exp\Big[ - 4 \ln(2) \frac{t^2}{\tau^2} \Big] \cos[t + g(t,\zeta) ] \label{las0}\ee
\be a_y(0,t) = \frac{a_0}{\sqrt{2}} \exp\Big[ - 4 \ln(2) \frac{t^2}{\tau^2} \Big] \sin[t + g(t,\zeta) ] \label{las1}\ee
\be g(t,\zeta) = \zeta \Big[ 4\ln(2) \frac{t^2}{\tau^2} + \frac{\tau^2}{16\ln(2) (1+\zeta^2)}\Big] + \frac{\tan^{-1}(\zeta)}{2} \label{las2} \ee  
here, $a_0$ is peak laser amplitude in dimensionless units, $\tau$ is dimensionless FWHM of the laser pulse, and $\zeta$ is the chirp parameter. The laser pulse profiles of unchirped, positively and negatively chirped are illustrated in Fig. \ref{pulse}. The 
time dependent frequency of the laser pulse is then given by \cite{Mackenroth2016_PRL},
\be \Omega(t) = 1 + \zeta \frac{8\ln(2)}{\tau^2} t \label{las3}\ee
For, $\zeta > 0$ ($\zeta < 0$) the low (high) frequency part interacts with the target first, followed by the high (low) frequency part.    

To understand how the chirp affects the transmission coefficient of the laser pulse, we have calculated the transmission coefficient for different chirp values by numerically solving the Eqs. \ref{wavepropa}, \ref{gama2} - \ref{las3}. The transmission coefficient of a positively, negatively and unchirped laser pulse ($a_0 = 0.5$, $\tau = 5$ cycles) for different target thickness with $n_0 = 1 n_c$ is compared, and the results are presented in Fig. \ref{waveTrans}. We observe from Fig. \ref{waveTrans}, that as we increase the target thickness the transmission coefficient for the unchirped pulse drops by $\sim 30\%$ with $\sim 100\%$ increase in the target thickness. However, for chirped pulses, the decrease in the transmission coefficient with target thickness is marginal ($\sim 5\%$) with the same variation in the target thickness. It can be understood by the nature of the chirped pulse itself. If the variation in the target thickness is smaller than the wavelength of the pulse, then, in that case, the transmission coefficient associated with the longer (smaller) wavelength (frequency) would not be affected. On the other hand for shorter (larger) wavelength (frequency) component, the skin depth is anyway much larger than the thickness of the target. The collective effect of the chirping would manifest in more or less similar transmission coefficients as we vary the target thickness in the sub-wavelength domain. In the following, we compare the results of this simplified wave propagation model with 1D PIC simulation.

\subsection{Comparison with PIC simulations}

Next, we turn to a comparison of the simplified wave propagation model (Fig. \ref{waveTrans}) with PIC-simulations. The 1D
Particle-In-Cell simulation (LPIC++) \cite{Lichters1997_MPQ} is carried out to study the
effect of target thickness on the transmission coefficient for different chirped values. We have modified this open-source 1D-3V PIC code,  to include the multilayer targets, chirped Gaussian laser pulses, and associated diagnostics. In this code the electric fields are normalized as we earlier discussed ($a_0 = eE/m_e\omega c$). However space and time
are taken in units of laser wavelength ($\lambda $) and one laser cycle $\tau =\lambda /c$ respectively, mass and charge are normalized with electron mass and charge respectively. We have used 100 cells per laser wavelength with each cell having 50 electron and ion macroparticles. The spatial grid size and temporal time step for the simulation are considered to be 0.01$\lambda$ and 0.01$\tau$ respectively. 

  \begin{figure}[t]
 \includegraphics[width=9cm,height=6cm]{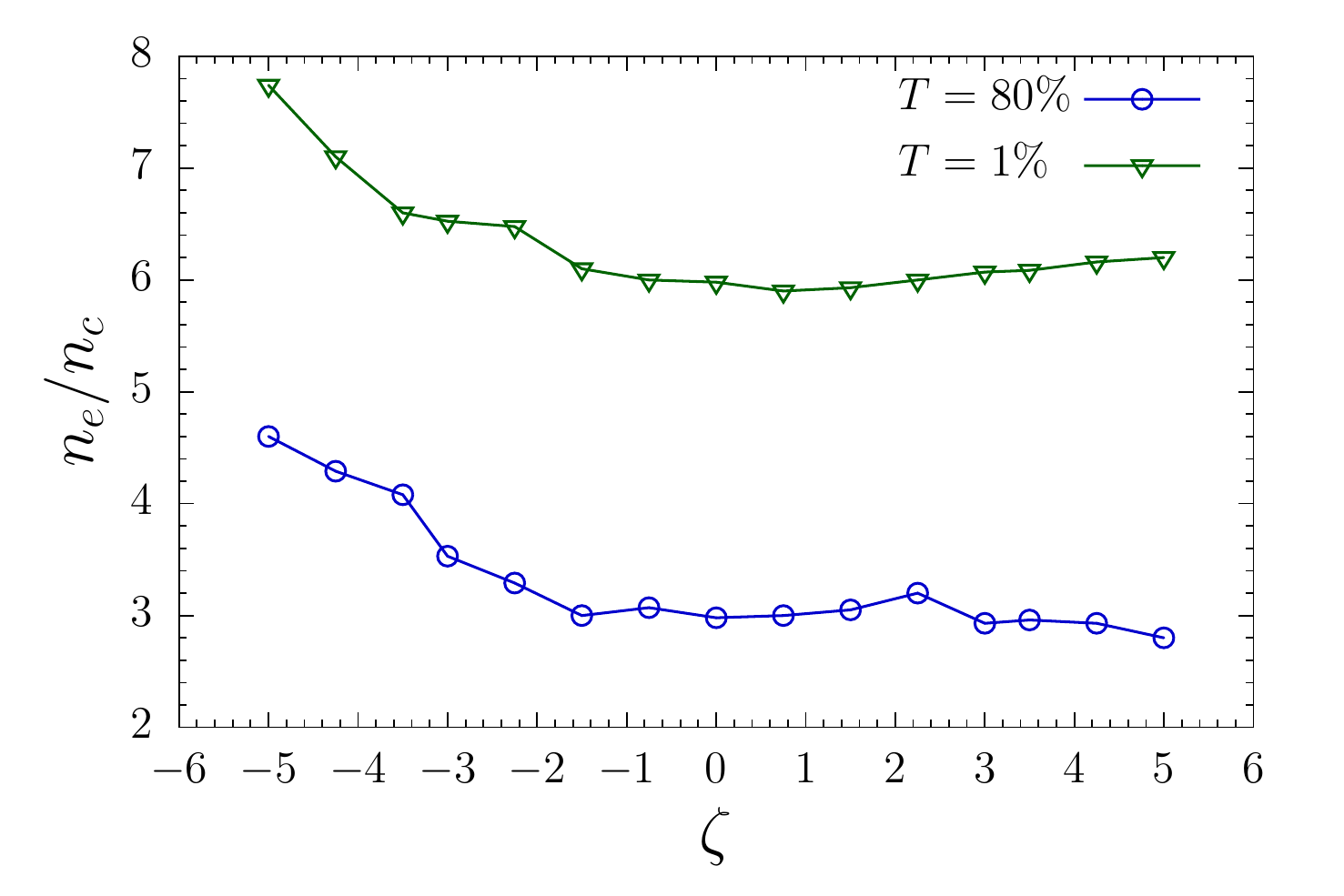}
 \caption{Variation of threshold target density for $80\%$ and $1\%$ transmission with chirp parameter. }
 \label{chirpVary}
\end{figure}

In Fig. \ref{lpicTrans}, we present the transmission coefficient dependence on the target thickness for different chirped values as calculated by the PIC simulations. The agreement with the simplified wave propagation model (Fig. \ref{waveTrans}) is found to be excellent. It is clear that for $a_0 = 0.5$, the electron heating is not very pronounced, or we would have observed the effect of the positive (low frequency interacts first) and negatively (high frequency interacts first) chirped pulses. The approximation we made in wave propagation model regarding the $p_z^e$, and $\Omega(t)$ are found to be consistent with the PIC simulations as well.

\begin{figure*}[t]
 \includegraphics[totalheight=10cm]{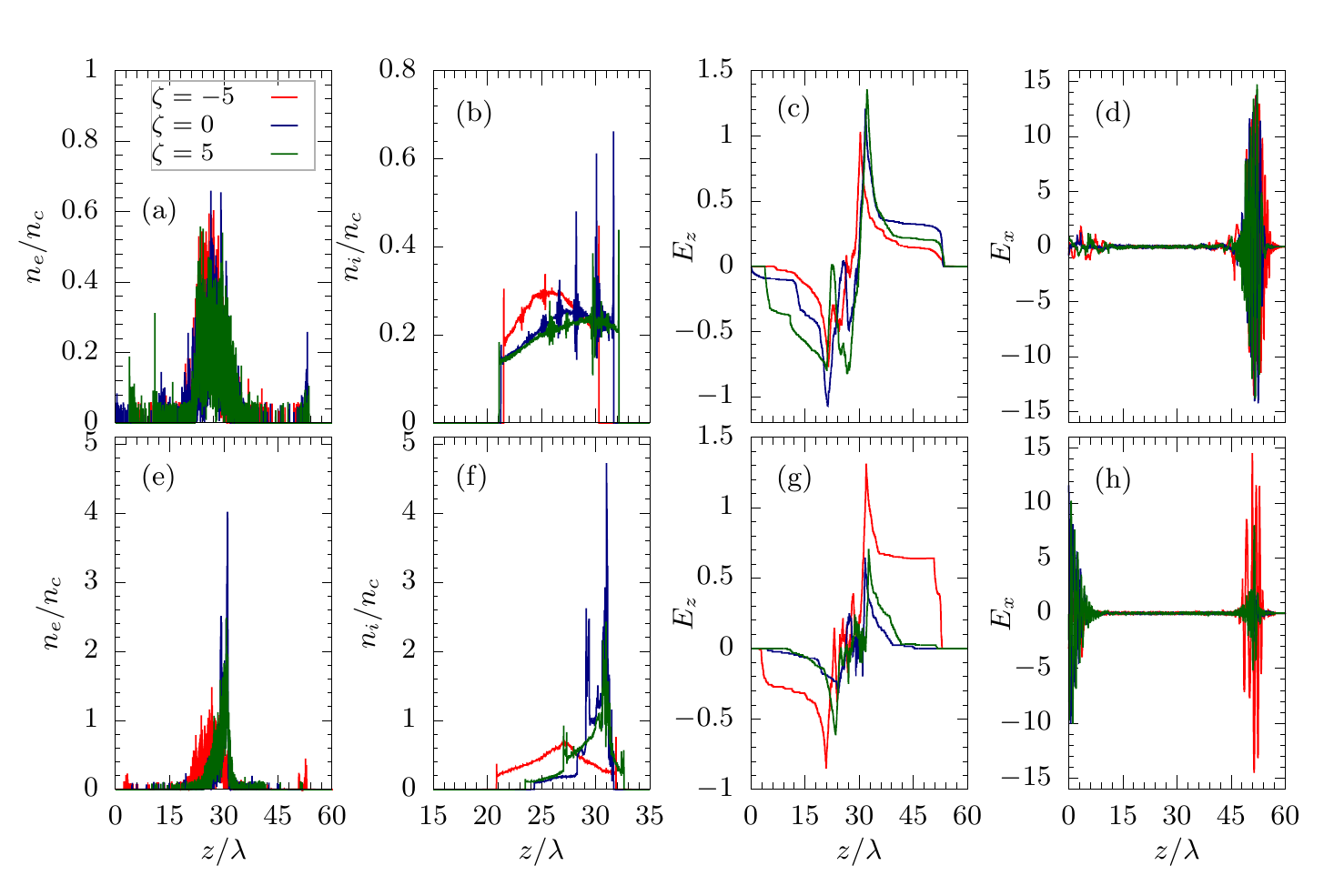}
 \caption{The effect of pulse chirp is illustrated for two different target densities, $3n_c$ (upper panel) and $6n_c$ (lower panel). 
 The spatial snapshot at $60\tau$ for the electron density (a,e), ion density (b,f), longitudinal electric field (c,g) and transverse laser 
 profile (d,h) is presented for $a_0 = 20$, $\tau = 5$ cycles and $d = 0.75\lambda$ (the target is placed at $25\lambda$) . }
 \label{geninfo}
\end{figure*}

\section{Results and discussions}
  
As we have seen, the chirp of the laser pulse can enhance the effective transmission of the laser pulse over unchirped laser pulses. For $a_0 = 0.5$ we did not observe a very prominent difference between the positively and the negatively chirped pulses. This is so, as the interaction dynamics is mostly governed by the transverse motion of the electrons. The omission of the $p_z^e$ in our simplified wave propagation model seems to be consistent with the fully relativistic 1D PIC simulation. However, for $a_0 \gg 1$, this might not be true, as the process is too non-linear to be approximated by this simple wave propagation model.  

The chirp effect on the transmission coefficient and on the interaction in general would be much pronounced for $a_0 \gg 1$, as the positively chirped pulse tends to compress the target initially, increasing the target density for the high-frequency part to interact. To study the interaction of the high intense ($a_0 \gg 1$) chirped laser beams with thin targets in RSIT regime, the kinetic simulations are essential, as the fluid model can no longer be used for such scenarios.

Now we study the interaction of the Gaussian, Circularly polarized laser pulse having peak amplitude $a_0 = 20$, and FWHM duration of 5 cycles with a target of thickness $0.75\lambda$. The 1D-3V PIC code LPIC++ is used for this purpose \cite{Lichters1997_MPQ}. The simulation domain is considered to be $100\lambda$ and the target (protons + electrons) of thickness $0.75\lambda$ is placed at $25\lambda$. The laser incidents on the target from the left side. It should be noted that for the cases when $a_0/[\pi (n_e/n_c) (d/\lambda)] < 1$, the ion acceleration is mainly dominated by the Ligh Sail mechanism \cite{macchi_ls,macchi_njp}. On the other hand, the RSIT mechanism begins to prevail in the regime where the ratio $a_0/[\pi (n_e/n_c) (d/\lambda)] \gtrsim 1$. The parameters used in the current study ($a_0 = 20$, $d = 0.75\lambda$ and $ n_e \lesssim 8 n_c$) clearly indicates that the RSIT regime would prevail. The laser parameters used i.e. $a_0 = 20$, Circularly Polarized are routinely accessible in ELI laser facility \cite{Sharma2018_SR}. 

\subsection{Chirp effect on threshold plasma density}

We present the variation of the threshold plasma density with chirp parameter ($\zeta$) in Fig. \ref{chirpVary}. Here, \lq\lq \emph{threshold plasma density}\rq\rq\ is the target density which allows some percentage fraction of the incident pulse to pass through the target for a given chirp parameter. The threshold plasma density for $80\%$ and $1\%$ transmission coefficients are presented for $-5 \leq \zeta \leq 5$ (see, Fig. \ref{chirpVary}). We observe that the threshold plasma density increases for negatively chirped laser pulses in either scenario ($80\%$ and $1\%$ transmission). In case of the positively chirped ($\zeta > 0$) pulses, the low-frequency part interacts with the target initially followed by the high-frequency component. The low-frequency component tends to compress the electron layer, increasing the electron density for the high-frequency component to interact. However, in case of negatively chirped pulses ($\zeta < 0$) the high-frequency component interact with the target initially followed by the low-frequency part. For high-frequency EM wave, corresponding critical density is also high, which enable it to transmit through the target without much of attenuation.

\begin{figure}[b]
 \includegraphics[width=0.5\textwidth]{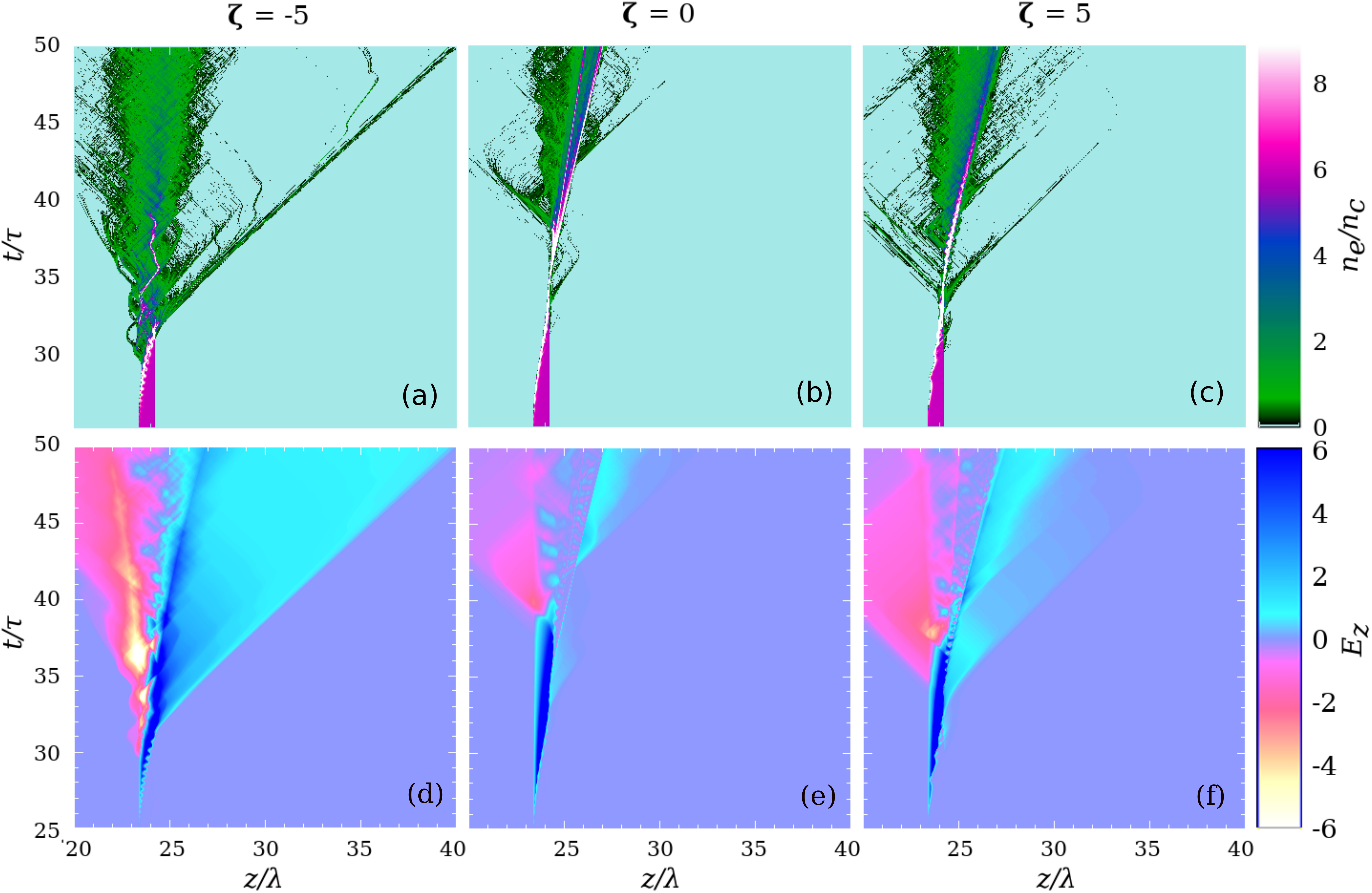}
 \caption{Spatio-temporal profile of the electron density (upper panel) and the longitudinal electrostatic field (lower panel) are presented for chirp parameters $\zeta = -5$ (left column), $\zeta = 0$ (center column) and $\zeta = 5$ (right column). The laser parameters are same as Fig. \ref{geninfo} with target density $n_e = 6n_c$. }
 \label{spacetime}
\end{figure}

In Fig. \ref{geninfo}, the spatial snapshots (as evaluated at $60\tau$) of the electron density, ion density, longitudinal field and laser field for different chirp values are shown. The snapshots are also compared for two different target densities, viz $3n_c$ (upper panel) and $6n_c$ (lower panel). We observe that for $3n_c$ case all the quantities are showing the similar characteristics for different chirp values. However, for $6n_c$ case we can see the distinctive spike of the electron density [see, Fig. \ref{geninfo}(e)] at $\sim 50 \lambda$  for $\zeta = -5$, this manifests in enhanced flat longitudinal electrostatic field [see, Fig. \ref{geninfo}(g)]. This can be 
understood from the fact that the transmission coefficient of the target with density $3n_c$ is $\gtrsim 80 \%$ for $-5 \leq \zeta \leq 5$ (see, Fig. \ref{chirpVary}). Moreover, for $6n_c$ case only the negatively chirped pulse will be having the $\gtrsim 1\%$ transmission (see, Fig. \ref{chirpVary}), in fact the pulse with $\zeta = -5$ can have $\sim 1\%$ transmission for the target with density $\sim 8n_c$. The transmission of the negatively chirped pulse can be observed in Fig. \ref{geninfo}(h). Furthermore, for $\zeta = 0$ and $\zeta = 5$, the transmission is $< 1\%$, as a consequence, the associated pondermotive force of the laser pulse tend to push the electrons inside the target, increasing the electron density [see, Fig. \ref{geninfo}(e)]. The electrostatic field formed by the compressed electron layer tend to pull the target ions, increasing the ion density as well [see, Fig. \ref{geninfo}(f)]. Now, on the contrary for $\zeta = -5$, the transmission coefficient is $> 1\%$, and hence as the pulse exits the target it drags the electrons with it as well. This motion of the electrons can be observed in the Fig. \ref{geninfo}(e), as a small spike in electron density around $\sim 50\lambda$ coincides with the location of the pulse after transmission, Fig. \ref{geninfo}(h). From the above analysis one can deduce that even $1\%$ transmission of intense laser beams is strong enough to heat the electrons to relativistic energies. As the electrons escape the target for $\zeta = -5$ case, it leaves the target positively charged, resulting in the expansion of the 
target ions in either direction, as seen in Fig. \ref{geninfo}(f).     

 \begin{figure*}[t]
\centering\includegraphics[totalheight=8.5cm,trim={1cm  1cm 1cm  1cm}]{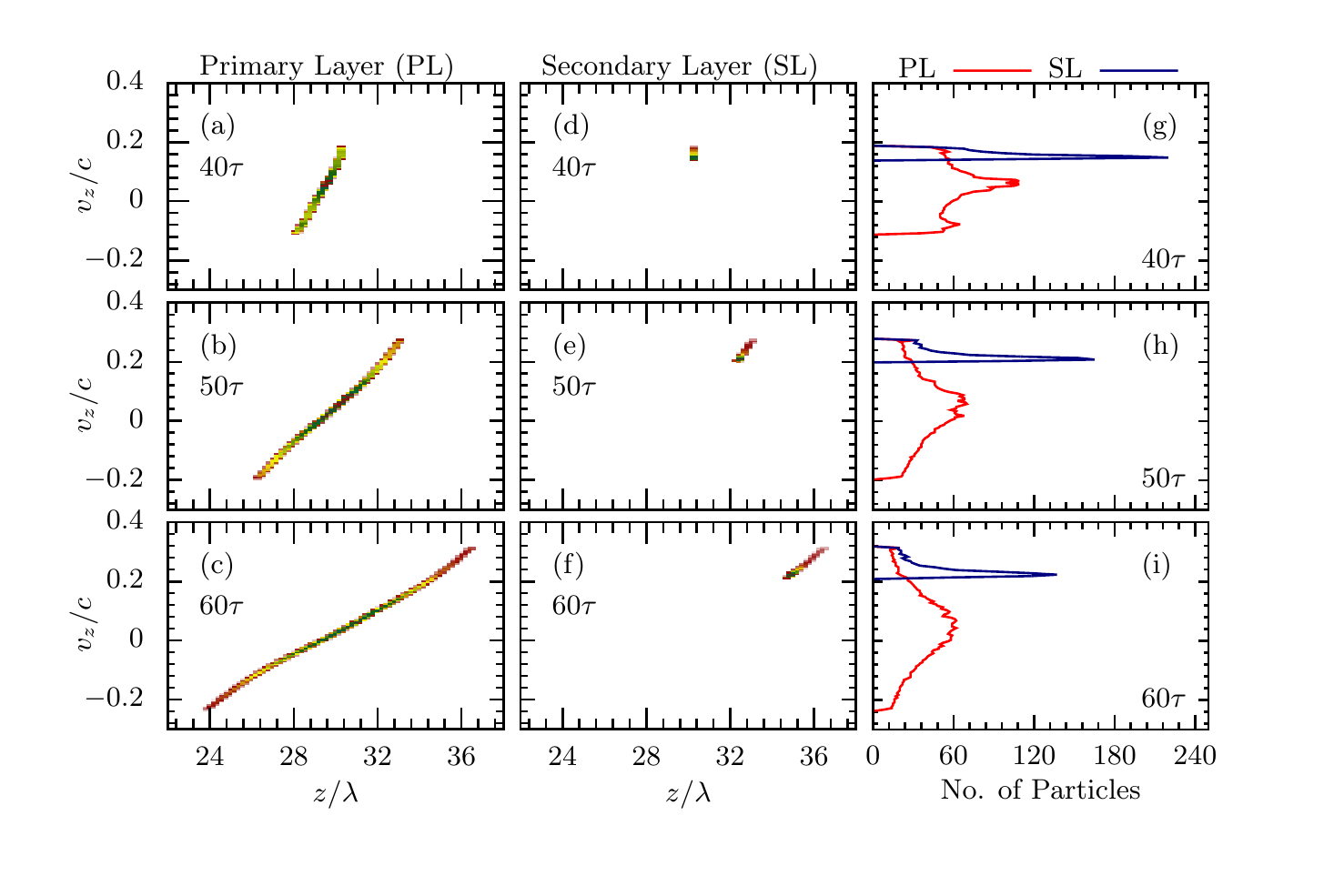}
\caption{Phasespace plots for the ions from the primary layer ($d = 0.75\lambda$, $n_{e} = 6n_c$) at different time instances (left column) and  ions from secondary layer ($0.2\lambda, 0.1n_c$) are presented (center column). The velocity spectrum for the ions from  primary and secondary layers are also illustrated (right column).}
 \label{phase}
\end{figure*}

\subsection{Spatio-temporal evolution of electrostatic field}
So far we have learned that the transmission of the pulse for $\zeta = -5$ results in efficient heating of the electrons followed by a very persistent electrostatic field, a few electrons are dragged away from the target. The rapid heating and excursion of the electrons from the target leaves the target charged, as a consequence the target ions expand under its own coulomb repulsion. To further elucidate this fact, in Fig. \ref{spacetime} we present the spatial and temporal evolution of the electron density and longitudinal electrostatic field for $\zeta = -5$, 0 and 5 for the case when target density is $6n_c$, all other laser parameters are same as Fig. \ref{geninfo}. The compression of the target can be seen for unchirped and positively chirped laser pulses. However, for the negatively chirped pulse ($\zeta = -5$), some electrons are accelerated by the transmitted pulse and starts co-moving with the laser pulse, this is mainly due to the nature of the circularly polarized pulse. For circularly polarized pulse the suppression of $\mb{J}\times\mb{B}$ heating of the electrons leads the push along the direction of the pulse propagation. This excursion of the electrons is the reason; we can see the approximately constant longitudianl electrostatic field configuration for $z \geq 25\lambda$ in Fig. \ref{spacetime}(d). We will see in the following how this kind of constant electrostatic field can be harnessed to obtain a mono-energetic proton bunch.

\subsection{Need for secondary layer}
We have seen so far, that the negatively chirped pulse efficiently create a very stable electrostatic field, as it transmits through the target. However, the excursion of the electrons leaves the target positively charged, as a consequence the target ions expand in either direction because of the Coulomb repulsion of the ions itself. The expansion of the target ions in either direction manifests in very broad energy distribution, on the contrary for any practical applications, a \textit{mono-energetic} ion bunches are desirable. To have a mono-energetic ion bunches from the current setup, a very thin, low density ($< n_c$) secondary layer is introduced just behind the primary layer. The low density of the secondary layer ensures that the interaction dynamics and in general the formation of the electrostatic field by the primary layer remains mostly unaffected even by the presence of the secondary layer. As the laser passes through this composite target (primary + secondary), the electrons are dragged away with the laser pulse, and the ions from the secondary layer experience a very persistent electrostatic field, leading to their acceleration as a mono-energetic ion bunch. 

 \begin{figure}[b]
 \includegraphics[totalheight=5.5cm]{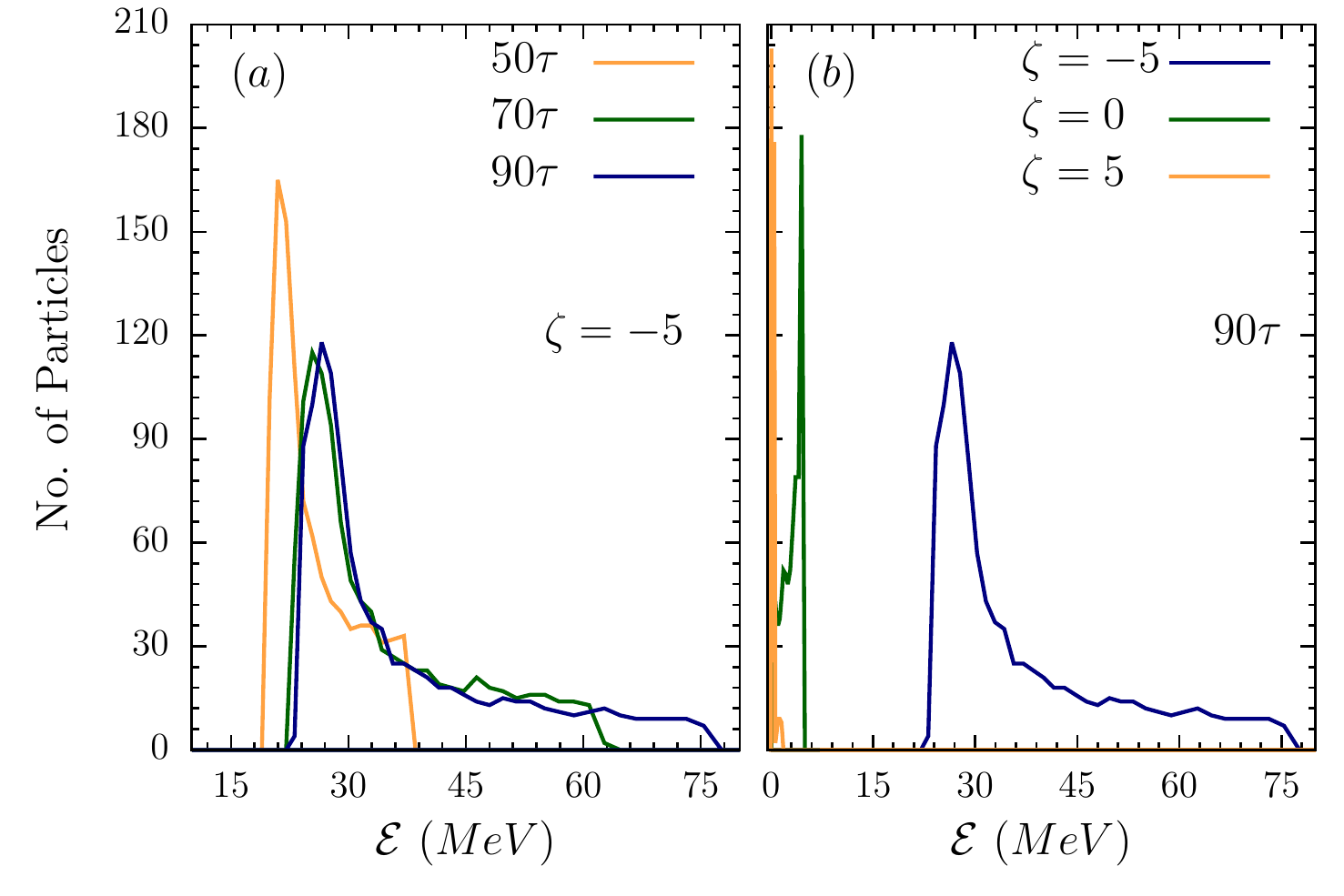}
 \caption{The energy spectrum of the ions from the secondary layer are presented at different time instances for $\zeta = -5$ (a). Moreover, the energy spectrum of secondary ions as evaluated at $90\tau$ for different chirp parameters is also compared (b). The laser parameters are same as Fig. \ref{phase}.}
 \label{enerspec}
\end{figure}

Next, we study the interaction of the negatively chirped ($\zeta = -5$), Gaussian, circularly polarized, 5 cycle laser pulse having peak amplitude $a_0 = 20$ with the composite target. The thickness (density) of the primary layer (PL) is considered to be $d = 0.75\lambda$ ($n_{e} = 6n_c$). However, for the secondary layer (SL) the thickness and density are considered to be $0.2\lambda$ and $0.1n_c$ respectively. In Fig. \ref{phase}, the phasespace plots of ions from PL and SL are presented at $40\tau$, $50\tau$ and $60\tau$. The expansion of the ions of the PL in either direction is visible in Fig. \ref{phase}(a,b,c). However, the ions from the SL are found to be accelerated as a bunch, Fig. \ref{phase}(d,e,f). The velocity spectrum of the ions from the PL and SL are also illustrated for different time instances in Fig. \ref{phase}(g,h,i). We present the energy spectrum of the ions from the SL in Fig. \ref{enerspec} at $50\tau$, $70\tau$ and $90\tau$. We observe that for the negatively chirped pulse with given laser and target parameters, the maximum number of ions from the SL are accelerated to $\sim 30$ MeV, however, the maximum energy ($\mathcal{E}_{max}$) of the bunch is observed to be $\sim 75$ MeV, as evaluated at $90\tau$. For the sake of completeness, the effect of the pulse chirp on the energy spectrum of the SL is also illustrated in Fig. \ref{enerspec}. For the positively and unchriped laser pulses the target density $6n_c$ will be in the opaque regime, as a consequence the laser will be reflected from the PL. The heating of the electrons at the rear side of the PL is not very efficient, and hence the ions from the SL are not very efficiently accelerated for the positively and unchirped laser pulses.

 \begin{figure}[b]
 \includegraphics[totalheight=6.cm,trim={1cm 0cm 1cm 0cm}]{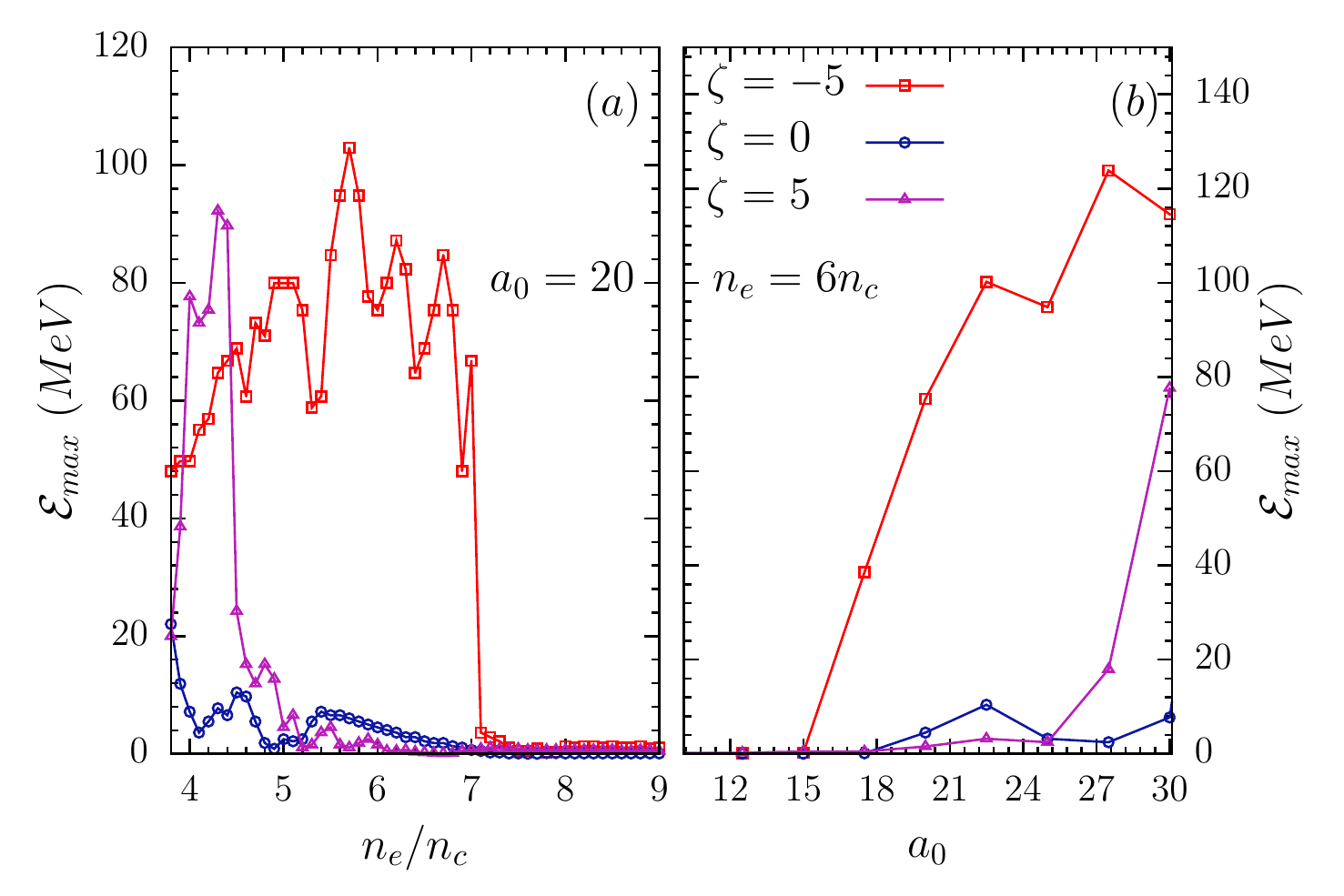}
 \caption{The effect of the pulse chirping on the maximum energy of the ions from the secondary layer is presented for different primary target density (a). The thickness of the primary layer is $0.75\lambda$ and peak laser amplitude is $a_0 =20$. The variation of the maximum ion energy with $a_0$ is presented in (b), here we have fixed the density of the primary layer to $6n_c$.}
 \label{pri_opt}
\end{figure}

\section{Optimization} 

In Fig. \ref{pri_opt} , the variation of the maximum energy of the ions from the SL is presented for different chirp parameters. 
The effect of the pulse chirp is illustrated for fixed laser amplitude $a_0 = 20$ and different primary target densities (a), and for fixed primary target density and different laser amplitudes (b). As the target density for $a_0 = 20$ is varied, we observe that for $\zeta = -5$, the maximum energy of $\sim 105$ MeV is obtained for $\sim 6n_c$. The threshold plasma density for $a_0 = 20$ and $\zeta = -5$ is $\sim 7n_c$ [see, Fig. \ref{chirpVary}] and hence the transmission of the pulse for $6n_c$ is $> 1\%$, leading towards the stable electrostatic formation as we discussed in the previous sections. However, for $\zeta = 5$ case, the target with density $6n_c$ would be opaque, and hence ions from SL will not be efficiently accelerated. We further observe that for $\zeta = 5$ the maximum ion energy is seen to be for the case when target density is $\sim 4 n_c$. As we discussed earlier, the leading low-frequency component of the positively chirped pulse tends to compress the target density by the radiation pressure, as a consequence the following high-frequency component interacts with the high-density target, resembling a similar scenario as negatively chirped pulse interacting with the high-density target. We have also studied the effect of the laser intensity on the maximum ion energy from the secondary layer for fixed target density of $6n_c$. It can be seen from Fig. \ref{pri_opt}(b) that for $a_0 \lesssim 15$, the target with density $6n_c$ and thickness $0.75\lambda$ would be opaque, and hence the electrostatic field generation is suppressed and so the acceleration of the ions from the secondary layer. However, as the $a_0$ increases an efficient acceleration is observed for the negatively chirped pulse.

\begin{figure}[t]
 \includegraphics[totalheight=6.cm,trim={1cm 0.5cm 1cm 0cm}]{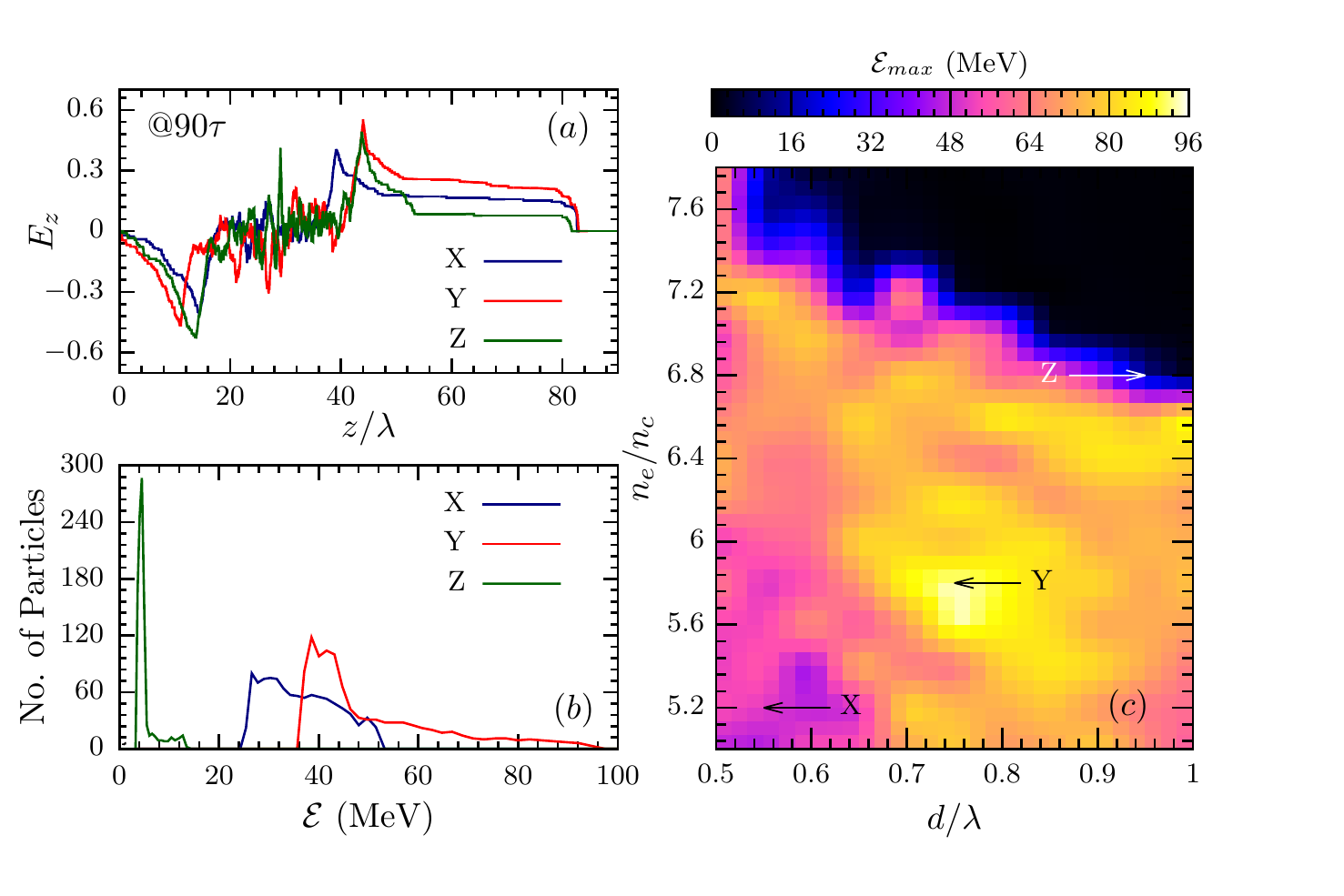}
 \caption{The longitudinal electrostatic field (a) and energy spectrum of the ions from secondary layer (b) are  presented for three 
 different primary target conditions. However, the maximum ion energy of the secondary layer is also presented for different thicknesses and densities of the primary layer (c). The laser pulse with peak amplitude $a_0 = 20$, duration 5 cycles and chirp parameter $\zeta = -5$ is considered. The target conditions ($d/\lambda,n_e/n_c$) are $X$ (0.55,5.2), $Y$ (0.75,5.8), and $Z$ (0.95,6.8). All the quantities are evaluated at $90\tau$. }
 \label{dtopt}
\end{figure}

So far we have seen that the negatively chirped pulses are efficient in generating very persistent and stable electrostatic field behind the target. The effect of the thickness and density of the PL on the maximum ion energy by the negatively chirped pulse is presented in Fig. \ref{dtopt}. We have varied the thickness of the PL from $0.5\lambda -  1\lambda$, and density from $5n_c - 8n_c$, and maximum ion energy (as evaluated at $90\tau$) of the ions from SL ($0.2\lambda, 0.1n_c$) are calculated for $a_0 = 20$, $\zeta = -5$, 5 cycles, circularly polarized, Gaussian laser pulse. In Fig. \ref{dtopt}, we have also presented the electrostatic field and energy spectrum of the ions from the secondary layer as evaluated at $90\tau$ for three different target parameters ($d/\lambda,n_e/n_c$) namely $X$ (0.55,5.2), $Y$ (0.75,5.8), and $Z$ (0.95,6.8). In all of the three cases we observe very stable flat electrostatic field behind the primary layer [see, Fig. \ref{dtopt}(a)]. The optimum target parameters for given laser conditions are found to be $Y$ (0.75,5.8) where maximum ion energy is observed to be $\sim 100$ MeV. We have fixed the parameters of the SL throughout the simulations. The main purpose of the SL is to have an accelerated proton bunch. The parameters of the SL neither alter the electrostatic field formed by the primary layer nor affects the laser pulse propagation, and hence the ions of SL mere serve as test particles.

\section{Concluding remarks}

In this work, we have studied the effect of the laser chirping on the acceleration of the protons via relativistic self-induced transparency. In the negatively (positively) chirped pulse, the high (low) frequency component of the pulse interacts with the target initially followed by the low (high) frequency component. The temporal variation of the frequency in the chirped EM pulse manifests in the associated time-dependent critical density, as a consequence the threshold plasma density of the negatively chirped laser pulse is found to be comparatively higher than the unchirped and positively chirped laser pulses. The initial low-frequency interaction of the positively chirped pulse with the target tend to compress the target layer by the radiation pressure, as a consequence, the target would be opaque for positively chirped pulses. Furthermore, as the negatively chirped pulse transmits through the target, the suppression of the $\mb{J}\times\mb{B}$ heating of the circularly polarized laser results in longitudinal push on the electrons; as a result, few electrons get dragged away and start co-moving with the laser pulse. This imbalance leaves the target positively charged followed by the expansion of the target ions under its own coulomb repulsion. However, the removal of the electrons also generate a very stable and persistent electrostatic field behind the primary layer which can be harnessed by the composite target geometry, comprised of  the low density, thin secondary layer behind the primary layer. The ions from the secondary layer are found to be accelerated as a bunch under the effect of the longitudinal field created by the primary layer (secondary layer does not affect the field formation in a profound manner) upon interaction by the negatively chirped laser pulse with this composite target. 

Under optimum conditions the maximum energy of the ions from the secondary layer is found to be $\sim 100$ MeV for circularly polarized, Gaussian, 5 cycles FWHM, negatively chirped ($\zeta = -5$), laser with peak amplitude $a_0 = 20$. These parameters for 800 nm laser  would translates to $\sim 10$ fs (intensity FWHM) pulse with peak intensity $\sim 8.5\times 10^{20}$ W/cm$^2$. However, similar energies are reported in the past, but with much higher $a_0$ values. For example, in Ref. \cite{Robinson2012_PPCF} the authors have demonstrated the ion acceleration in the HB-RPA regime and reported the proton energies $\sim 150$ MeV by irradiating a laser with peak amplitude $a_0 \sim 90$. 
Similarly, under the RPA regime, ion acceleration to $\sim$ 150 MeV with $a_0 = 108$ has also been reported in Ref. \cite{Macchi2010_NI}.

In summary, we have studied the effect of the pulse chirp on the transmission of the laser pulse through the sub-wavelength target and 
associated ion acceleration. The transmission coefficient for $a_0 = 0.5$ are estimated by a simplified wave propagation model which 
takes into account the time dependent critical density of the target. The results of this simplified wave propagation model are found to be
 consistent with the 1D fully relativistic PIC model. In this work we have used the chirp model which is beyond the linear approximation. The chirp  model used in this study is in close analogy of the idea of the Chirped Pulse Amplification. Furthermore, we studied the interaction of the intense laser pulse with $a_0 = 20$ with the target having the thickness $0.75\lambda$ and density $6n_c$ for different chirp parameters. It has  been observed that the negatively chirped laser pulse is very efficient in creating a stable and persistent electrostatic field behind 
 the target. The electrostatic field created behind the target can be harnessed by a low density, thin secondary layer behind the target. 
The optimization of the target parameters are finally carried out to have a maximum energy of the accelerated ions of the secondary layer.  

The feasibility of the proposed scheme under experimental scenario needs full 3D Particle-in-Cell simulations, which is currently beyond the scope of the current manuscript. Soon, we plan to study the effect of the higher dimensions on the ion acceleration under RSIT regime. For this purpose we would like to explore the 3D PIC codes like {\small\sc Epoch} \cite{epoch}, or {\small\sc Piccante} \cite{piccante}.        

\section*{Acknowledgments}
Authors would like to acknowledge the Department of Physics, Birla Institute of Technology and Science, Pilani, Rajasthan, India for the 
computational support.

\end{document}